\documentclass[aps,prl,showpacs,amsmath,amssymb,twocolumn,groupedaddress]{revtex4-1}
 \usepackage{graphicx}
\usepackage{dcolumn}
\usepackage{verbatim}
\usepackage{bm}
\begin{document}
 \pagestyle{plain}
\title{Radii and Binding Energies in Oxygen Isotopes: A Challenge for Nuclear Forces} 
\author{V.~Lapoux$^{1}$}\email[E-mail address: ]{vlapoux@cea.fr}
\author{V.~Som\`a$^{1}$}
\author{C.~Barbieri$^{2}$}
\author{H.~Hergert$^{3}$}
\author{J. D.~Holt$^{4}$}
\author{S. R.~Stroberg$^{4}$}
\affiliation{$^1$ CEA, Centre de Saclay, IRFU, Service de Physique Nucl\'eaire, 91191 Gif-sur-Yvette, France}
\affiliation{$^2$  Department of Physics, University of Surrey, Guildford GU2 7XH, United Kingdom} 
\affiliation{$^3$ National Superconducting Cyclotron Laboratory and Department of Physics and Astronomy,
Michigan State University, East Lansing, Michigan 48824, USA}
\affiliation{$^4$ TRIUMF, 4004 Wesbrook Mall, Vancouver, British Columbia, V6T 2A3, Canada}

\date{\today}
\begin{abstract}
We present a systematic study of both nuclear radii and binding energies in 
(even) oxygen isotopes from the valley of stability to the neutron drip line. 
Both charge and matter radii are compared to state-of-the-art {\it ab initio} 
calculations along with binding energy systematics. Experimental matter
radii are obtained through a complete evaluation of the available elastic 
proton scattering data of oxygen isotopes. We show that, in spite of a good 
reproduction of binding energies, {\it ab initio} calculations with 
conventional nuclear interactions derived within chiral effective field theory 
fail to provide a realistic description of charge and matter radii. A novel 
version of two- and three-nucleon forces leads to considerable 
improvement of the simultaneous description of the three observables for 
stable isotopes but shows deficiencies for the most neutron-rich systems. 
Thus, crucial challenges related to the development of nuclear interactions remain.
\end{abstract}

\maketitle
 
Our present understanding of atomic nuclei faces the following major questions. Experimentally, we aim 
(i) to determine the location of the proton and neutron 
drip lines~\cite{Erler12, Thoennessen2015}, i.e. the limits in 
  neutron   numbers $N$ upon which, for fixed proton number $Z$,
  with decreasing or increasing   $N$, nuclei are not bound with
respect to particle emission, and
(ii) to measure nuclear structure observables offering systematic tests of microscopic models.
While nuclear masses have been experimentally determined for the majority of known light
and medium-mass nuclei~\cite{AME12},
measurements of charge and matter radii are typically more challenging.
Charge radii for stable isotopes have been accessed in the past by means of electron
 scattering~\cite{Hof56}.
In recent years, laser spectroscopy experiments allow extending such measurements to 
unstable nuclei with lifetimes down to a few milliseconds~\cite{Laser95}.
Matter radii are determined by scattering with hadronic
probes which requires a modelization of the reaction mechanism.
Theoretically, intensive works have also been performed also towards linking a universal
description of atomic nuclei to elementary interactions~\cite{Machleidt01,QCDab,EFTth11}
amongst  constituent nucleons and, ultimately, to the underlying  theory of strong interactions,
quantum chromoDynamics (QCD).
If accomplished, this {\it ab initio} description would be beneficial
both for a deep understanding of known nuclei (stable and unstable, totalling around 3300)
and to predict on reliable bases
the features of undiscovered ones (few more thousands are expected). 
Many of the latter are not, in the foreseeable future,
experimentally at reach, yet they are crucial  to understanding nucleosynthesis 
phenomena, modelled using large sets of evaluated data and  of calculated  observables.

The reliability of first-principles calculations depends upon a consistent understanding of fundamental  
observables:  ground-state characteristics of nuclei related to their existence (masses, 
expressed as binding energies) and sizes (expressed as root mean square -rms- radii).
Special interest resides in the study of masses and sizes for a given element along isotopic chains. 
Experimentally, their determination is increasingly difficult as one approaches the neutron drip line;
as of today, the heaviest element with available data on all existing bound isotopes is oxygen 
($Z$=8)~\cite{AME12}. 
Using theoretical simulations, the link between nuclear properties and inter-nucleon forces can be explored 
for different {\it N/Z} values,
thus, critically testing both our knowledge of nuclear forces and many-body theories.

In this Letter, we focus on oxygen isotopes for which,  
in spite of the tremendous progress of recent {\it ab initio} methods, a simultaneous reproduction 
of masses and radii has not yet been achieved.
We present important findings from novel {\it ab initio} calculations along with a complete
evaluation of matter radii, $r_m$, for stable and  neutron-rich oxygen isotopes. 
Here, $r_m$  are deduced via a microscopic reanalysis of proton elastic scattering data 
sets. They complement charge radii $r_{\text{ch}}$, offering an extended comparison through
the isotopic chain that allows testing state-of-the-art many-body calculations.
We show that a recent version of two- and three-nucleon (2$N$ and 3$N$) forces 
leads to considerable improvement in the  critical description of radii.

A viable \textit{ab initio} strategy consists  in exploiting the separation of scales between
QCD and (low-energy) nuclear dynamics, taking point nucleons as degrees of freedom. 
For decades, realistic $2N$ interactions were built from fitting scattering data,
 see, {\it e.g.},~\cite{Machleidt01}. 
However, model limitations were seen through discrepancies with experimental data,
like underbinding of finite nuclei and inadequate saturation properties 
of extended nuclear matter. More recently, the approach consisted in using the principles of 
chiral effective field theory (EFT) to provide a systematic construction
 of nuclear forces,  a well-founded starting point for structure calculations~\cite{QCDab, EFTth11}.
Many-body techniques have, themselves,  undergone major progress and extended their domain of 
applicability both in mass and in terms of accessible (open-shell) isotopes for a given 
element~\cite{Soma11, Her13, Som13, Bin14, Her14, Holt13a, Holt14, Hag14, Bog14, Jan14, Sig15, Dug15}. 
An emblematic case that has received considerable attention  is oxygen binding energies, where several 
calculations have established the crucial role played by 3$N$ forces in the
reproduction of the neutron drip line at $^{24}$O~\cite{Ots10, Hag12o, Holt13b, Her13, Cip13, Lah14, Heb15}.
The excellent agreement between experimental data and  calculations 
based on a next-to-next-to-next-to-leading order (N$^3$LO) 2$N$ 
and N$^2$LO 3$N$ chiral interaction by Entem, Machleidt and others (EM)~\cite{EM03, Nav07, Roth12} was
greeted as a milestone for {\it ab initio} methods, even though a consistent description of nuclear radii could not be
achieved at the same time~\cite{Cipollone15}.
Since then, this deficiency has remained a puzzle.
Subsequent calculations of heavier systems~\cite{Som13, Bin14, Her14} and infinite 
nuclear matter~\cite{Car13, Hag14nm} confirmed the systematic 
underestimation of charge radii, a sizable overbinding and
too spread-out spectra, all pointing to an incorrect reproduction of the 
saturation properties of nuclear matter.
While interactions with good saturation 
properties existed~\cite{Heb11,Cor14,Simo16}, this problem led to the 
focused development of a novel nuclear interaction,  NNLO$_{\text{sat}}\,$~\cite{NNLOeks15},
which includes contributions up to N$^2$LO in the chiral EFT expansion (both in the
2$N$ and 3$N$ sectors) and differs from EM in two main aspects. 
First, the optimization of the (``low-energy") coupling constants is performed 
simultaneously for 2$N$ and 3$N$ terms~\cite{Car15}; EM, in contrast, optimizes 3$N$ forces subsequently. 
Second,  in addition to observables from few-body ($A$=2,3,4) systems, experimental constraints from light nuclei 
(energies and charge radii in some C and O isotopes) are  included in the optimization.
This  aspect departs from the  strategy of   EM, in which parameters 
in the $A$-body sector are fixed uniquely by observables in $A$-body systems.
Although first applications point to good predictive power
for ground-state properties~\cite{NNLOeks15,Hag15,Ruiz16}, the performance of the NNLO$_{\text{sat}}\,$ potential remains 
to be tested along complete isotopic chains.

Here, we employ two different many-body approaches, self-consistent Green's function
(SCGF) and in-medium similarity renormalization group (IMSRG),
each available in two versions. The first are based on standard expansion schemes and, thus, applicable
only to closed-shell nuclei (e.g., not $^{18,20}$O): Dyson SCGF (DGF)~\cite{Dic04} and single-reference IMSRG (SR-IMSRG)~\cite{Tsu11}
respectively. The second are built on Bogoliubov-type reference states and thus allow for a proper treatment
of pairing correlations and  systems displaying an open-shell character. 
These are labeled Gorkov SCGF (GGF)~\cite{Soma11} and multireference 
IMSRG (MR-IMSRG)~\cite{Her13}, respectively.
For the MR-IMSRG, the reference state is first projected on good proton and neutron numbers.
Having different {\it ab initio} approaches at hand is crucial for benchmarking theoretical results
and inferring as unbiased as possible information on the input  forces.
Moreover, while DGF, 
SR-IMSRG and MR-IMSRG feature a comparable content in terms of many-body expansion,
GGF currently includes a lower amount of many-body correlations, which allows testing the many-body 
convergence~\cite{Som13}.

\begin{figure}       
             \includegraphics[width=8.0cm]{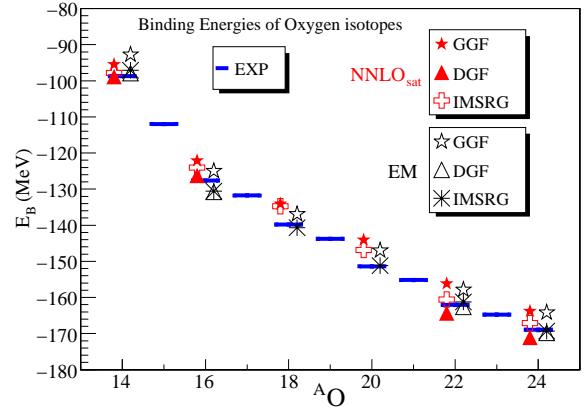} 
  \caption{Oxygen binding energies. Results from SCGF (DGF and GGF) and IMSRG
  calculations  with
  EM  and NNLO$_{\text{sat}}\,$   
  are displayed along with   experimental data.}
   \label{fig:plotEbExpTheoOxyNNLO}
\end{figure}

First, we compute  binding energies $E_B$ for $^{14-24}$O for the two sets of 2$N$ and 3$N$ 
interactions with the four many-body schemes. EM is further evolved to a low-momentum
scale $\lambda=1.88 - 2.0\,\text{fm}^{-1}$ by means of SRG techniques~\cite{Bog07,Bog10}.
Results are displayed in Fig.~\ref{fig:plotEbExpTheoOxyNNLO}. For both interactions, different many-body calculations 
yield values of $E_B$ spanning intervals of up to 10 MeV, from 5 to 10$\%$ of the total. 
 Compared to experimental binding energies, EM and NNLO$_{\text{sat}}\,$ perform similarly, following the
trend of available data along the chain both in absolute and in relative terms.
Overall, results shown in Fig.~\ref{fig:plotEbExpTheoOxyNNLO}
confirm previous findings for EM and validate the use along the isotopic chain for NNLO$_{\text{sat}}$.
 
Now, we examine the nuclear charge observables. In addition to $r_{\text{ch}}$ radii, analytical forms of 
fitted experimental charge densities can be extracted from ($e$,$e$) cross sections. 
Standard forms include two- or three-parameter Fermi (2PF or 3PF) profiles~\cite{Vries87}. 
By unfolding~\cite{SatLove79} the finite size
of proton charge distribution [whose $r_{\text{ch}}$ radius is 0.877(7) fm~\cite{PDG2010}],
proton ground-state  densities $\rho_p$ can be deduced, and the corresponding  $r_p$ radius
defined as the rms radius of the $\rho_p(r)$  distribution ($\sqrt{\langle r^2 \rangle}$).
It should be underlined that, due to the various analysis techniques providing
charge densities, the global systematic error  on $r_p$  is significantly larger (roughly $0.05$ fm) than 
the one on single $r_{\text{ch}}$ values (of the order of 0.01 fm).
For $^{16}$O, $r_{\text{ch}}$ was
estimated to be 2.730 (25) fm~\cite{Sick70} and 2.737 (8) fm~\cite{Miska79,Vries87}.
Differences in $r_{\text{ch}}$ between $^{17,18}$O and $^{16}$O,
$\Delta r_{\text{ch}}~=- 0.008 (7)$ and   $+0.074  (8)$  fm~\cite{Miska79},
are affected by the same systematic errors.
 
In this Letter, we determine  matter radii via the proton probe. 
We consider angular distributions of proton elastic scattering cross sections
and compare data to calculations performed  using a microscopic density-dependent optical model potential (OMP)
inserted in the distorted wave Born approximation (DWBA). Recently, this type of analysis 
has been  successfully applied to the case of helium isotopes, for which $r_m$ radii were
extracted with uncertainties of the order of 0.1 fm~\cite{EPJArmsHe}.
We employ the energy- and density-dependent Jeukenne-Lejeune-Mahaux (JLM) potential~\cite{JLM77b}, derived from 
a $G$-matrix formalism and extensively tested in the analysis of 
nucleon scattering data for a wide range of nuclei.
This complex potential depends only on the incident energy $E$ and on 
neutron and proton densities. Here, we use the standard form:\\
$U_{\text{JLM}} (\rho,E)  = \lambda_V V(\rho,E)  + i \lambda_W W(\rho,E) $,
with $\lambda_V=\lambda_W=1$.\\
For $^{18-22}$O,  nucleon separation energies are sufficiently high  to exclude strong 
coupling effects to continuum or to excited states, and the imaginary part is enough to include,
 implicitly, all other relevant coupled-channel effects.

For the stable symmetric $^{16}$O, $r_m$ was extracted from combined ($e$,$e$), ($p$,$p$) and ($n$,$n$) in Ref.~\cite{Pet85}
using the following procedure:  the (3PF) density profile $\rho_p$ was deduced
from electron scattering data~\cite{Sick70}, the same profile was assumed for the neutron 
density distribution. This ``experimental" matter density built from the ($e$,$e$) data was 
used to compute the potentials. 
This procedure was also followed for $^{17,18}$O, with the neutron density profiles initially 
taken as $(N/Z)*\rho_p$
then adjusted to reproduce elastic data  on heavy ions~\cite{SatLove79}.
We refer to  densities extracted in this way
as the experimental  (exp) ones, with $r_p$ values for $^{16-18}$O
given in Table~\ref{tab:ExpRmsEeSigpp}. 
\begin{figure}
       \centering{\includegraphics[width=4.2cm]{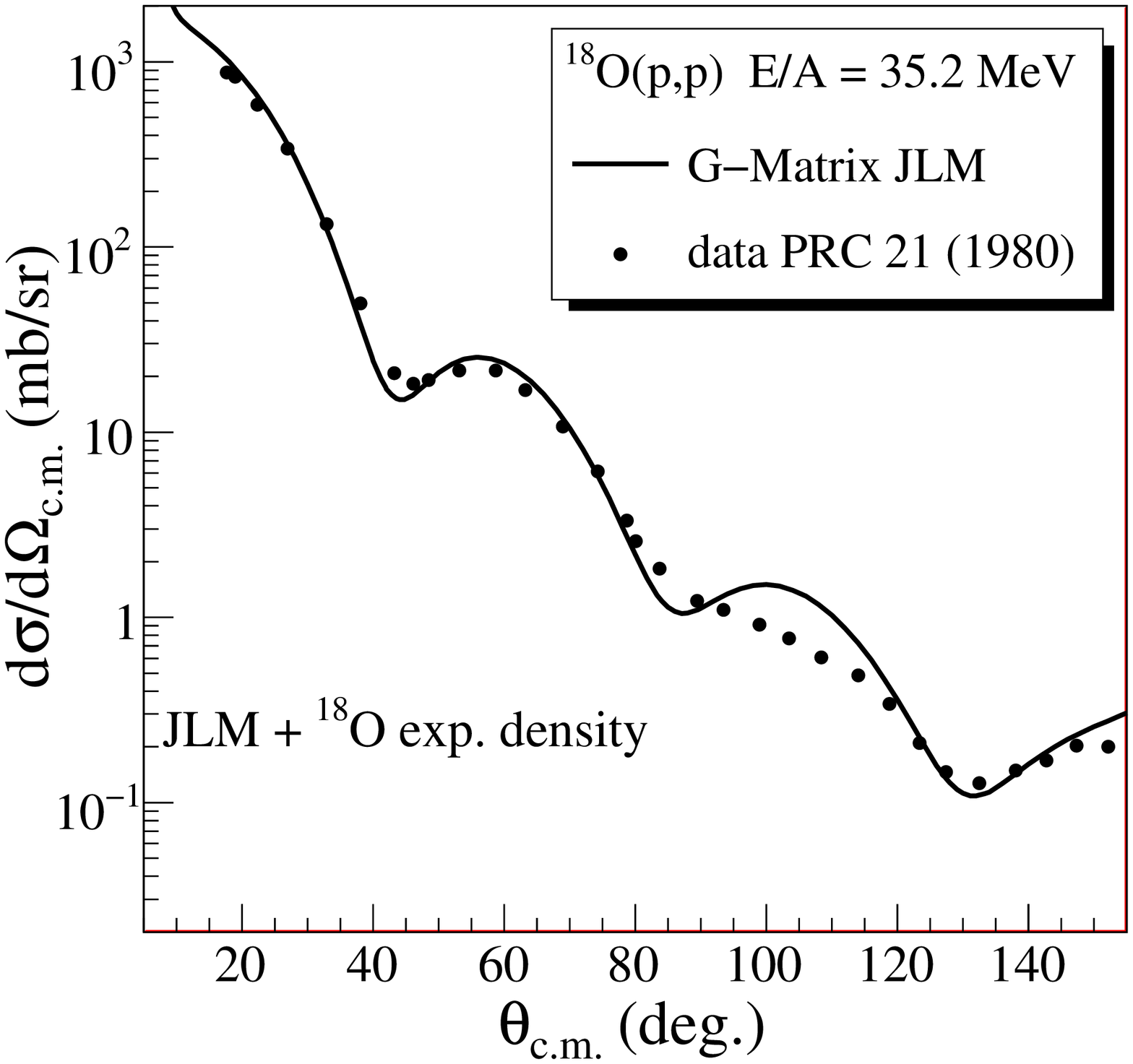}
 \includegraphics[width=4.2cm]{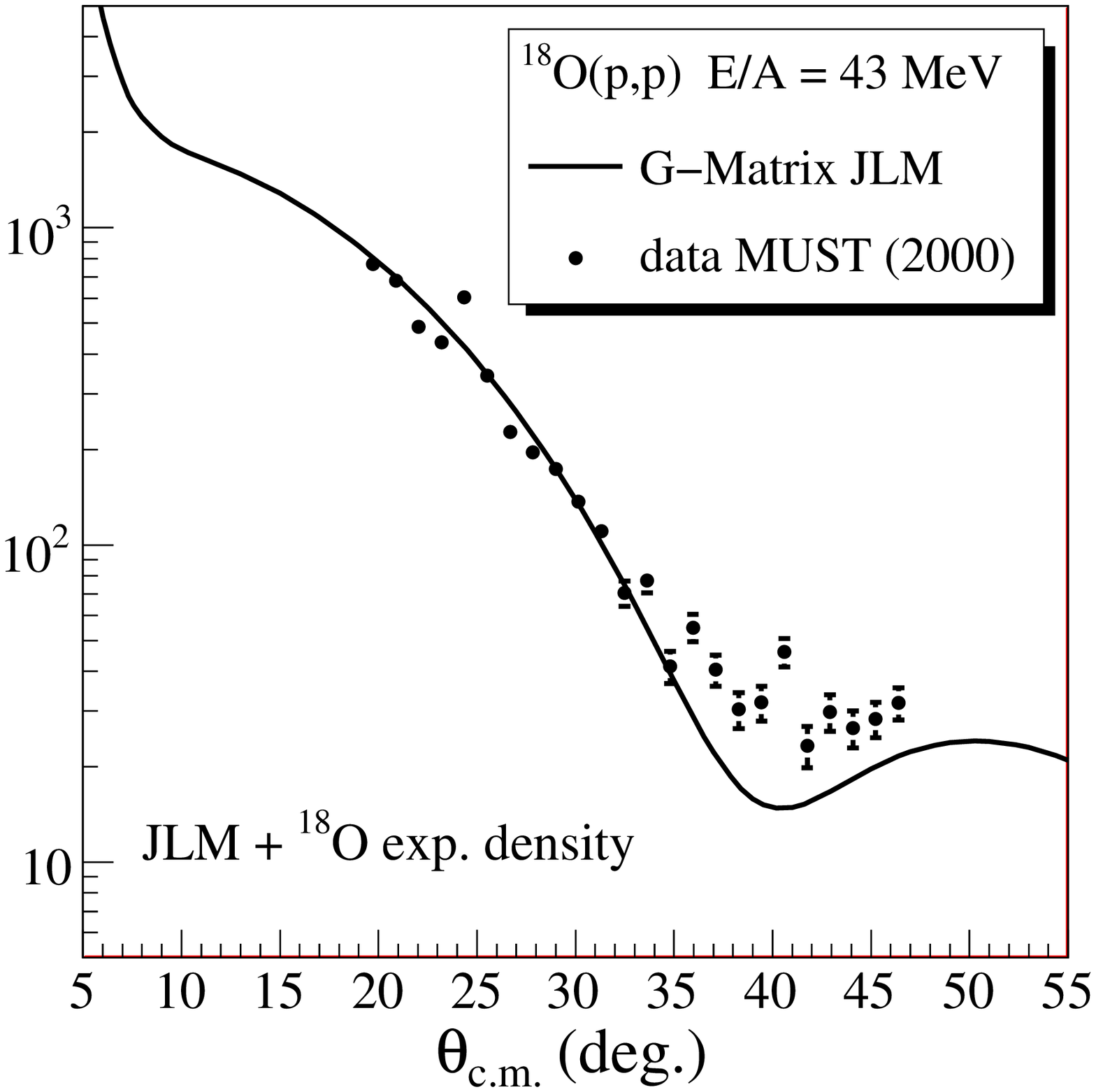}}
        \centering{\includegraphics[width=4.2cm]{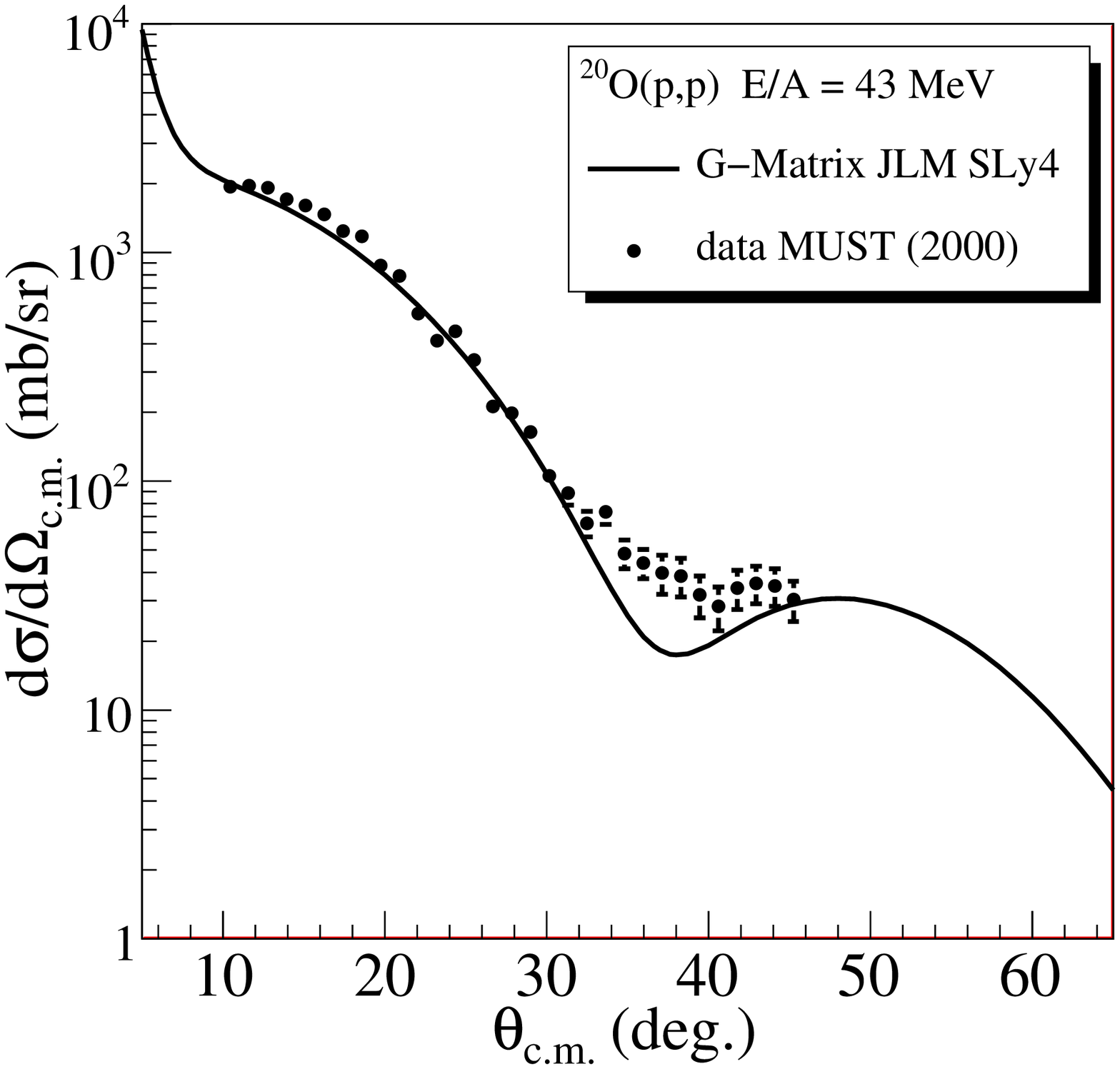}
        \includegraphics[width=4.2cm]{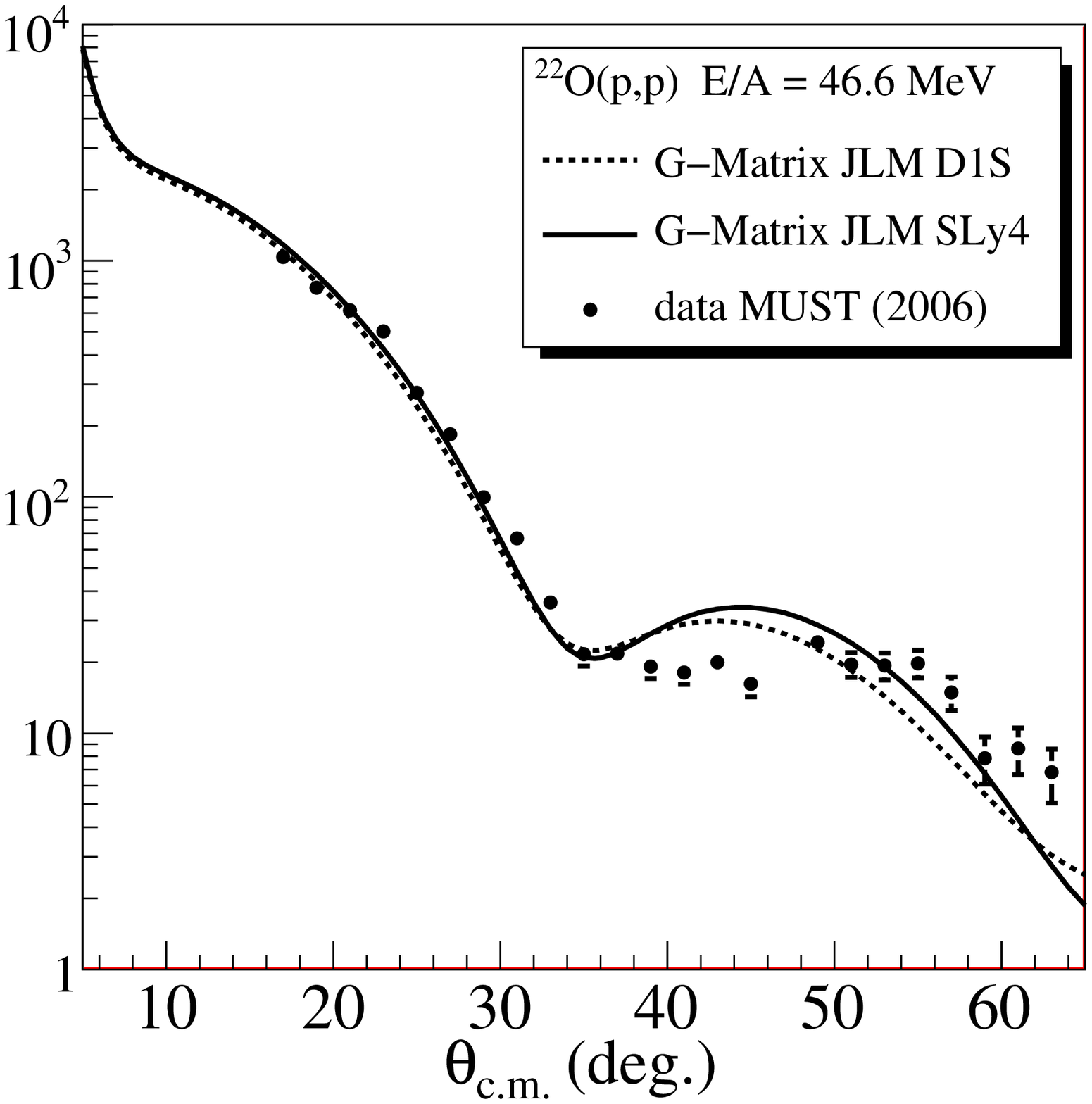}}
  \caption{Experimental elastic ($p$,$p$) distributions compared to OMP calculations ({\it this work}).
  (Top) $^{18}$O   (data:~\cite{Fab80, Khan00}).
  (Bottom) $^{20,22}$O (data:~\cite{Khan00,Bec06}).
  } 
   \label{fig:elasticOpp}
   \end{figure}
   
We first performed OMP calculations for $^{18}$O and compared them
to data collected at 35.2 $A\cdot$MeV in direct kinematics~\cite{Fab80} and at 43 $A\cdot$MeV
in inverse kinematics~\cite{Khan00}.
Starting from a 2PF profile fitted to exp densities, by changing the two parameters
governing size and diffusiveness,
we generated a family of densities then inserted into the OMP and fitted to   data.
Since only the most forward angles have small global errors and are sensitive to the size of the nucleus, 
we limited our fit to 46$^{\circ}$ and 33$^{\circ}$ for 35.2  and 43 $A\cdot$MeV data, respectively,
i.e., to data with statistical + systematic errors below 10\%.
For these degrees of freedom (DOF), by keeping the curves falling within $\chi^2 / \text{DOF} <1$, we determined an
associated matter radius $r_m = 2.75 (10)$ fm. 
The 2PF profiles with the same $r_m$ lead to very similar $\chi^2 / \text{DOF}$,
signaling that calculations, in the region of forward angles, are rather insensitive to the diffusiveness.
As shown in Fig.~\ref{fig:elasticOpp}, calculations are in good agreement
with ($p$,$p$) data, which confirms the validity of the OMP approach provided that realistic densities are 
employed.
We repeated the analysis using  densities generated 
by Hartree-Fock BCS calculations~\cite{Khan00} with Skyrme interactions, each associated with a different $r_m$.
Results are very similar to the ones of Fig.~\ref{fig:elasticOpp}, 
with   $r_m = 2.77 (10)$ fm, close to the one from `exp densities.
This validates the use of OMP calculations  
to estimate $r_m$ radii from ($p$,$p$) cross sections~\cite{EPJArmsHe}. 
 
For unstable $^{20,22}$O,   elastic proton scattering cross sections were measured using oxygen
beams at 43 and 46.6 $A\cdot$MeV, respectively~\cite{Khan00,Bec06}.
We performed   OMP  calculations with microscopic densities for $^{20,22}$O.
 Angular distributions up to 30$^{\circ}$ (for $^{20}$O) and 33$^{\circ}$ (for $^{22}$O) 
were considered for the fits.
Results are displayed in  Fig.~\ref{fig:elasticOpp}.
In order to show the sensitivity to  the microscopic inputs, 
we compare, for $^{22}$O,  results with densities from the Sly4~\cite{Sly4}  Skyrme interaction
with those obtained with densities from Hartree-Fock-Bogoliubov
calculations based on the Gogny  D1S force~\cite{DeGo80,D1S91}.
In both cases,  ($p$,$p$) cross sections are well reproduced.
Resulting  $r_m$  radii are 2.90~fm in $^{20}$O
along with 2.96 and 3.03~fm in $^{22}$O for Sly4 and D1S densities, respectively.
The sensitivity study led us to the same range of $\pm 0.1$~fm, which
is the uncertainty on our values throughout the  ($p$,$p$) analysis.
The results  are summarized in Table~\ref{tab:ExpRmsEeSigpp}.

\begin{table}[b]
\begin{tabular}{llll|ll} \hline\noalign{\smallskip}
A  & 16   &  17   & 18 & 20 & 22 \\
\hline\noalign{\smallskip}
 $r_p$            &       2.59  (7)   &    2.60 (8) &    2.68  (10)  &  & \\ 
 $r_m$ ($\sigma_I$)   &       2.54 (2)   &    2.59  (5) &    2.61  (8)  &  2.69(3)  & 2.88(6)   \\ 
$r_m$ ($p$,$p$)           &       2.60 (8)   &    2.67  (10) &    2.77  (10) & 2.9 (1)  & 3.0 (1) \\
                        \noalign{\smallskip}\hline
\end{tabular}
\caption{Experimental rms radii (in fm) of O isotopes:
  $r_p$  for $^{16-18}$O are extracted from charge densities~\cite{Vries87,SatLove79,Norum82}. 
For $A$ = 16,   $r_m$ is  evaluated from ($p$,$p$) data~\cite{Pet85}, and for $A$ = 17, 
  via heavy-ion scattering~\cite{SatLove79}. 
$r_m$ from $\sigma_I$ are given in Ref.~\cite{Oza01}. For $A$ =18-22,  ``$r_m$ ($p$,$p$)"
 values are from the present work and are explained in the text.}
\label{tab:ExpRmsEeSigpp}
\end{table}

\begin{figure}
          \includegraphics[width=6cm]{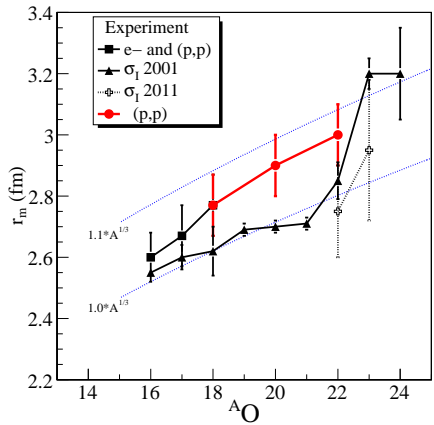}  
 \caption{Experimental values for the $r_m$ radii, deduced from $\sigma_I$, ($e$,$e$) and 
($p$,$p$) measurements (see Table~\ref{tab:ExpRmsEeSigpp}). Blue lines show the $A^{1/3}$ behavior
of the liquid  drop model.   
   }
   \label{fig:plotRmsExpTheoOxySigIpp}
\end{figure}

Studying interaction cross sections ($\sigma_I$)~\cite{Oza01} 
is another way of  deducing matter radii. In Fig.~\ref{fig:plotRmsExpTheoOxySigIpp}, 
we compare experimental $r_m$ radii for $^{16-22}$O from ($e$,$e$) and ($p$,$p$) to values
obtained from $\sigma_I$ measurements~\cite{Oza01, Kan11} (see, also, Table~\ref{tab:ExpRmsEeSigpp}).
While ($e$,$e$) and ($p$,$p$)   provide a consistent set of $r_p$ and $r_m$ radii
for $^{16-18}$O, this is not the case for $r_m$ values obtained from $\sigma_I$,
usually extracted without including correlations 
in the target, which arguably influences scattering amplitudes.
Since our analysis of the stable isotopes, used as a reference, 
provides $r_m$ radii with an uncertainty  of the order of 0.1 fm,
we also conclude that uncertainties deduced from $\sigma_I$ are underestimated.
Consequently, we focus on results obtained from ($e$,$e$) and ($p$,$p$) data
for the comparison with theory.

\begin{figure}      
         \centering{\includegraphics[width=7cm]{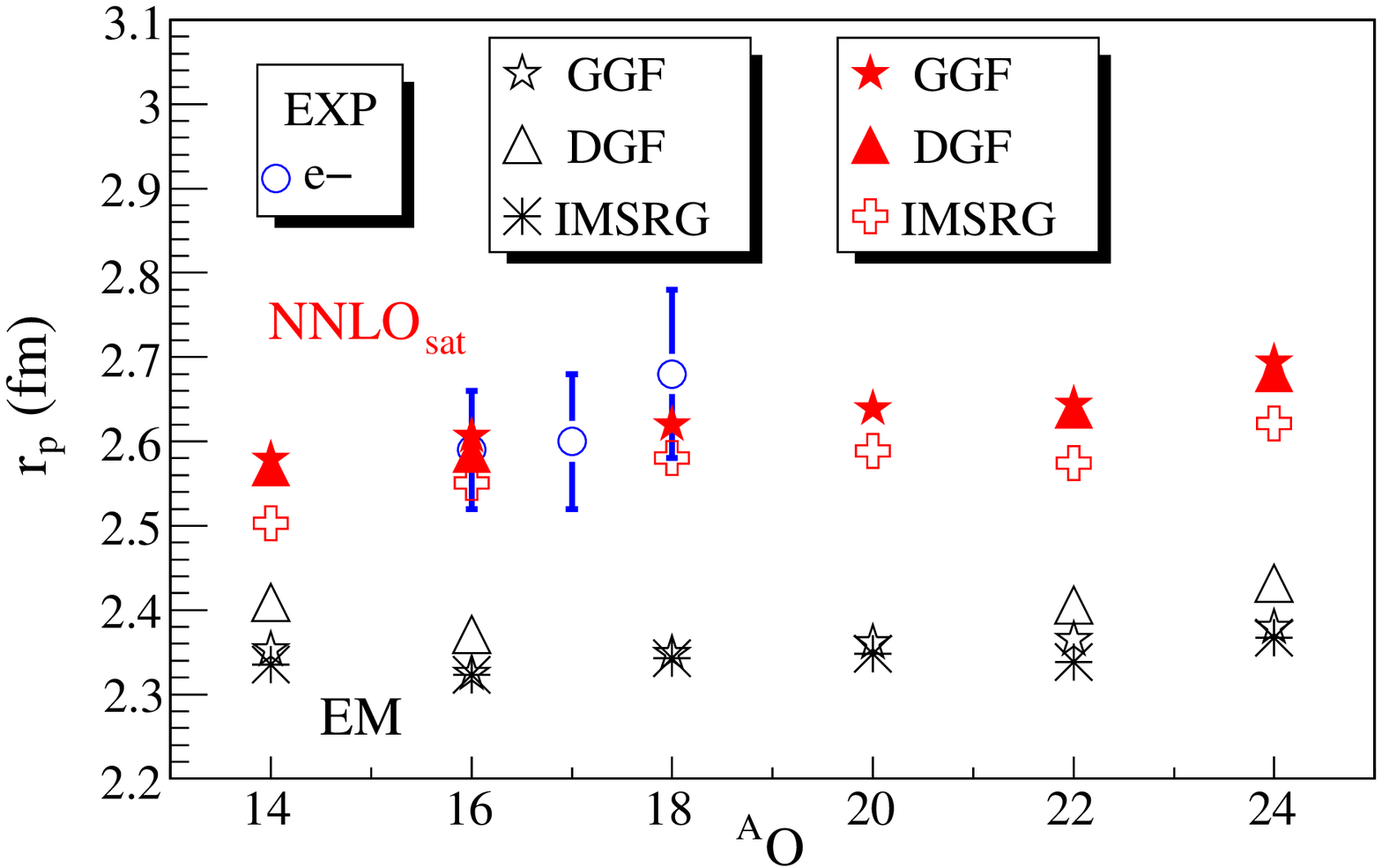}
              \includegraphics[width=7cm]{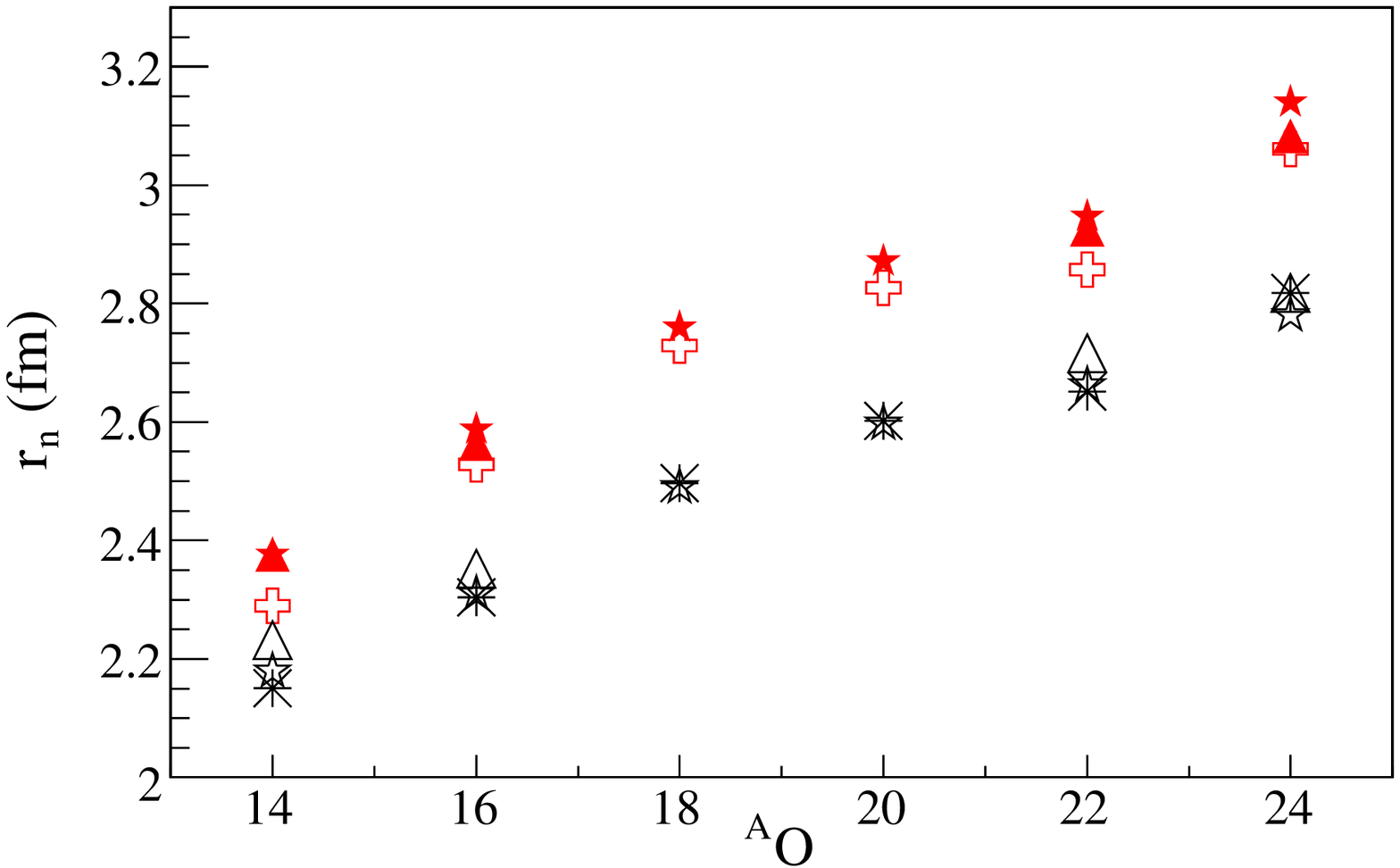}}
  \caption{Proton (top) and neutron (bottom)
  radii obtained from IMSRG and SCGF calculations
  with EM  and NNLO$_{\text{sat}}$.
 Experimental $r_p$ values are given in Table~\ref{tab:ExpRmsEeSigpp}.}
   \label{fig:plotRmsExpTheoOxyNNLOpn}
\end{figure}

\begin{figure}[t]
             \includegraphics[width=7cm]{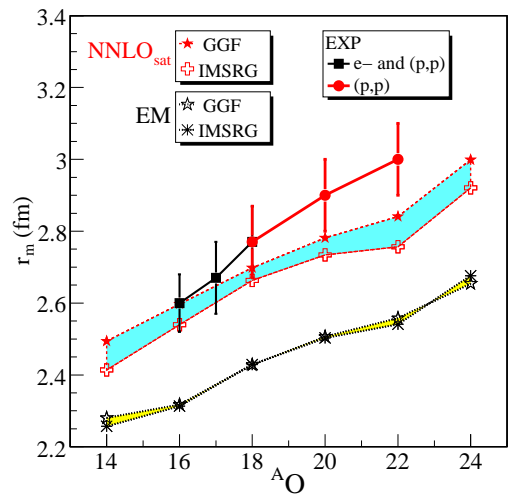} 
  \caption{Matter radii from our analysis and given in 
Tab.~\ref{tab:ExpRmsEeSigpp},   compared to   calculations with
  EM~\cite{EM03, Nav07, Roth12} and NNLO$_{\text{sat}}\,$~\cite{NNLOeks15}.
  Bands span results from GGF and MR-IMSRG   schemes. 
  }
   \label{fig:plotRmsExpTheoOxyNNLO}
\end{figure}

We start by analyzing  calculations for proton and neutron radii,
shown  in Fig.~\ref{fig:plotRmsExpTheoOxyNNLOpn}.
We notice that, for each interaction, there is good agreement between the various methods, which
span 0.05 (0.1) fm when EM (NNLO$_{\text{sat}}$) is used. 
This shows that different state-of-the-art schemes achieve, for a given interaction, 
an uncertainty that is smaller than (i) experimental uncertainty
and (ii) the uncertainty coming from the use of different interactions.
 Clear discrepancies are observed between radii computed with EM and NNLO$_{\text{sat}}$, with the former
being systematically smaller by 0.2-0.3 fm.
While EM largely underestimates data,  $r_p$ values are well reproduced by NNLO$_{\text{sat}}$,
  keeping in mind  that $r_{\text{ch}}$ of $^{16}$O is included in the NNLO$_{\text{sat}}\,$ fit.
The performance of the interactions along the isotopic chain can be seen for matter radii, where in
Fig.~\ref{fig:plotRmsExpTheoOxyNNLO} the evaluations from the ($p$,$p$) analysis
are compared to GGF and MR-IMSRG.
Similar conclusions are drawn by considering other  schemes,  e.g., see
Fig.~\ref{fig:plotRmsExpTheoOxyNNLOpn}, where rms radii computed with EM underestimate
evaluated data by about 0.3 - 0.4 fm for all isotopes. 
 
Results significantly improve with NNLO$_{\text{sat}}$, although the description deteriorates
towards the neutron drip line, with a discrepancy of about 0.2 fm in $^{22}$O. 
Recently, a similar effect was  observed for the calcium isotopes~\cite{Ruiz16}.

These results reinforce the progress 
of nuclear {\it ab initio} calculations, which  are able to address systematics
of isotopic chains beyond light systems and, thus, provide  critical feedback
on the long-term developments of   internucleon interactions.
To this extent, joint theory-experiment  analyses are essential
and have   to start with a realistic description of both sizes and masses.
In this work we focused on the oxygen chain, the heaviest one for which 
experimental information on both $E_B$ and radii is available up to the neutron drip line.
We showed that  nuclear sizes of unstable isotopes can be obtained through the ($p$,$p$) data 
analysis   within 0.1 fm. The combined comparison of measured charge-matter radii and $E_B$
with   {\it ab initio} calculations offers a unique insight 
on nuclear forces:
the current standard EM yields an excellent reproduction of $E_B$ but significantly underestimates
 radii, whereas the unconventional NNLO$_{\text{sat}}$ clearly improves the description of  radii.
Our results raise  questions about the choice of 
observables that should be included in the fit and the resulting predictive power whenever 
this strategy is followed.

More precise information on oxygen radii, e.g., $r_{\text{ch}}$ via laser spectroscopy measurements, 
would allow confirming our ($p$,$p$) analysis and further refining the present discussion.
Similar studies in heavier isotopes will also  contribute to
the systematic development of nuclear forces.
Finally, we stress that a simultaneous reproduction of binding energies and radii
in stable and neutron-rich nuclei is mandatory for reliable structure but even more for reaction
calculations. Scattering amplitudes and nucleon-nucleus interactions evolve as
a function of the size, which should be consistently taken into account 
 when more microscopic reaction approaches are considered.

\section*{acknowledgements}
The {\it Espace de Structure et 
de r\'eactions Nucl\'eaires Th\'eorique}  ESNT (http://esnt.cea.fr) 
 framework at CEA   is gratefully
acknowledged for supporting the project that initiated the present work.
The authors would like to thank T. Duguet for useful discussions and
P. Navr{\'a}til, A. Calci, S. Binder, J. Langhammer, and R. Roth
for providing the interaction matrix elements
used in the present calculations.
C. B. is funded by the Science and Technology Facilities Council (STFC) under Grant No. ST/L005743/1.
SCGF calculations were performed by using HPC resources from 
GENCI-TGCC (Contracts No. 2015-057392 and No. 2016-057392)
and the DiRAC Data Analytic system at the University of Cambridge 
(under BIS National E-infrastructure capital Grant No. ST/J005673/1,
and STFC Grants No. ST/H008586/1 and No. ST/K00333X/1).
H. H. acknowledges support by the NSCL/FRIB Laboratory. 
TRIUMF receives federal funding via a contribution agreement with the National
Research Council of Canada.
Computing resources for MR-IMSRG calculations
were provided by the Ohio Supercomputing Center (OSC) and the National Energy Research Scientific Computing 
Center (NERSC), a DOE Office of Science User Facility supported by the Office of Science of the U.S. Department 
of Energy under Contract No. DE-AC02-05CH11231.


\begin{references}
\bibitem{Erler12} J. Erler, N. Birge, M. Kortelainen, W. Nazarewicz,
E. Olsen, A. M. Perhac and M. Stoitsov, {\it The limits of the nuclear landscape},
 Nature {\bf 486}, 509 (2012).
\bibitem{Thoennessen2015} M. Thoennessen, Int. J. Mod. Phys. E {\bf 24}, 1530002 (2015).
\bibitem{AME12} G. Audi, M. Wang, A.H. Wapstra, F.G. Kondev, M. MacCormick, X. Xu, and B. Pfeiffer,
Chin. Phys. C {\bf 36},  1287 (2012). 
\bibitem{Hof56} R. Hofstadter, Rev. Mod. Phys. {\bf 28}, 214 (1956).
\bibitem{Laser95} J. Billowes and P. Campbell, J. Phys. G {\bf 21}, 707 (1995).
\bibitem{Machleidt01} R. Machleidt and I. Slaus,
 J. Phys. G {\bf 27}, (2001) R 69 (topical review), 
{\it and  references therein}. 
\bibitem{QCDab} E. Epelbaum, H.-W. Hammer, and U.-G. Mei{\ss}ner, Rev. Mod. Phys. {\bf 81}, 1773 (2009). 
\bibitem{EFTth11} R. Machleidt and D. Entem, Phys. Rep. {\bf 503}, 1 (2011).
\bibitem{Soma11} V. Som\`a, T. Duguet, and C. Barbieri, Phys. Rev. C {\bf 84}, 064317 (2011). 
\bibitem{Her13} H. Hergert, S. Binder, A. Calci, J. Langhammer, and
R. Roth, Phys.  Rev. Lett. {\bf 110}, 242501 (2013). 
\bibitem{Som13} V. Som\`a, A. Cipollone, C. Barbieri, P. Navr\'atil, and T. Duguet, 
Phys. Rev. C {\bf 89}, 061301 (2014). 
\bibitem{Bin14} S. Binder, J. Langhammer, A. Calci, and R. Roth, Phys.
Lett. B {\bf 736}, 119 (2014).
\bibitem{Her14} H. Hergert, S. K. Bogner, T. D. Morris, S. Binder,
A. Calci, J. Langhammer, and R. Roth, Phys. Rev. C
{\bf 90}, 041302 (2014).
\bibitem{Holt13a} J. D. Holt, J. Men\'endez, and A. Schwenk, Phys. Rev.
Lett. {\bf 110}, 022502 (2013).
\bibitem{Holt14} J. D. Holt, J. Menendez, J. Simonis, and A. Schwenk,
Phys. Rev. C {\bf 90}, 024312 (2014).
\bibitem{Hag14} G. Hagen, T. Papenbrock, M. Hjorth-Jensen, and D. J.
Dean, Rep. Prog. Phys. {\bf 77}, 096302 (2014).
\bibitem{Bog14} S. K. Bogner, H. Hergert, J. D. Holt, A. Schwenk,
S. Binder, A. Calci, J. Langhammer, and R. Roth, Phys.
Rev. Lett. {\bf 113}, 142501 (2014).
\bibitem{Jan14} G. R. Jansen, J. Engel, G. Hagen, P. Navr\'atil, and
A. Signoracci, Phys. Rev. Lett. {\bf 113}, 142502 (2014).
\bibitem{Sig15} A. Signoracci, T. Duguet, G. Hagen, and G. R. Jansen,
Phys. Rev. C {\bf 91}, 064320 (2015).
\bibitem{Dug15} T. Duguet, J. Phys. G {\bf 42}, 025107 (2015).
\bibitem{Ots10} T. Otsuka, T. Suzuki, J. D. Holt, A. Schwenk, and
Y. Akaishi, Phys. Rev. Lett. {\bf 105}, 032501 (2010).
\bibitem{Hag12o} G. Hagen, M. Hjorth-Jensen, G. R. Jansen, R. Machleidt, and  T. Papenbrock, 
Phys. Rev. Lett. {\bf 108}, 242501 (2012).
\bibitem{Holt13b} J. D. Holt, J. Men\'endez, and A. Schwenk, Eur. Phys. J.
A {\bf 49}, 39 (2013).
\bibitem{Cip13} A. Cipollone, C. Barbieri, and P. Navr\'atil,
 Phys. Rev. Lett. {\bf 111}, 062501 (2013).
\bibitem{Lah14}  T. A. L\"ahde, E. Epelbaum, H. Krebs, D. Lee, U.-G.
Mei{\ss}ner, and G. Rupak, Phys. Lett. B {\bf 732}, 110 (2014).
\bibitem{Heb15} K. Hebeler, J. D. Holt, J. Men\'endez, and A. Schwenk,
Annu. Rev. Nucl. Part. Sci. {\bf 65}, 457 (2015).
\bibitem{EM03} D.R. Entem and R. Machleidt, Phys. Rev. C {\bf 68}, 041001 (2003).
\bibitem{Nav07} P. Navr\'atil, Few-Body Syst. {\bf 41}, 117 (2007).
\bibitem{Roth12} R. Roth, S. Binder, K. Vobig, A. Calci, J. Langhammer,
and P. Navr\'atil, Phys. Rev. Lett. {\bf 109}, 052501 (2012).
\bibitem{Cipollone15} A. Cipollone, C. Barbieri, and P. Navr\'atil, Phys. Rev.
C {\bf 92}, 014306 (2015).
\bibitem{Car13} A. Carbone, A. Polls, and A. Rios, Phys. Rev. C {\bf 88}, 044302 (2013).
\bibitem{Hag14nm} G. Hagen, T. Papenbrock, A. Ekstr{\"o}m, K. A. Wendt,
G. Baardsen, S. Gandolfi, M. Hjorth-Jensen, and C. J. Horowitz, Phys. Rev. C {\bf 89}, 014319 (2014).
\bibitem{Heb11} K. Hebeler, S. K. Bogner, R. J. Furnstahl, A. Nogga, and
 A. Schwenk, Phys. Rev. C {\bf 83}, 031301(R) (2011).
\bibitem{Cor14} L. Coraggio, J. W. Holt, N. Itaco, R. Machleidt, L. E. Marcucci, 
and F. Sammarruca, Phys. Rev. C { \bf 89}, 044321 (2014).
\bibitem{Simo16} J. Simonis, K. Hebeler, J. D. Holt, J. Men\'endez, and A. Schwenk, 
Phys. Rev. C {\bf 93}, 011302(R) (2016).
\bibitem{NNLOeks15} A. Ekstr\"{o}m, G. R. Jansen, K. A. Wendt, G. Hagen,
T. Papenbrock, B. D. Carlsson, C. Forss\'en, M. Hjorth-Jensen, P. Navr\'atil, 
and W. Nazarewicz, Phys. Rev. C {\bf 91}, 051301 (2015).
\bibitem{Car15} B. D. Carlsson, A. Ekstr{\"o}m, C. Forss\'en, D. F. Str{\"o}mberg, G. R. Jansen,
O. Lilja, M. Lindby, B. A. Mattsson, and K. A. Wendt, Phys. Rev. X {\bf 6}, 011019 (2016).
\bibitem{Hag15} G. Hagen {\it et al.}, Nat. Phys. {\bf 12}, 186 (2015).
\bibitem{Ruiz16} R. F. Garcia Ruiz {\it et al.}, Nat. Physics  {\bf 12}, 594 (2016).
\bibitem{Dic04} W. H. Dickhoff and C. Barbieri, Prog. Part. Nucl. Phys.
{\bf 52}, 377 (2004).
\bibitem{Tsu11} K. Tsukiyama, S. K. Bogner, and A. Schwenk, Phys. Rev.
Lett. {\bf 106}, 222502 (2011).
\bibitem{Bog07} S. K. Bogner, R. J. Furnstahl, and R. J. Perry, Phys.
Rev. C {\bf 75}, 061001(R) (2007).
\bibitem{Bog10} S. K. Bogner, R. J. Furnstahl, and A. Schwenk, Prog.
Part. Nucl. Phys. {\bf 65}, 94 (2010).
 \bibitem{Vries87} H. De Vries, C. W. De Jager, and C. De Vries,  
At. Data Nucl. Data Tables {\bf 36}, 495 (1987), {\it and references therein}.
\bibitem{SatLove79} G. R. Satchler and W. G. Love, Phys. Rep. {\bf 55}, 183 (1979).
\bibitem{PDG2010} K. Nakamura {\it et al.}, (Particle Data Group), J. Phys. G {\bf 37}, 075021 (2010). 
\bibitem{Sick70} I. Sick and J. S. McCarthy, Nucl. Phys. {\bf A150}, 631 (1970).
\bibitem{Miska79} H. Miska, B. Norum, M. V. Hynes, W. Bertozzi, S. Kowalski, F. N. Rad, 
C. P. Sargent, T. Sasanuma, and B. L. Berman, Phys. Lett. {\bf 83}B, 165 (1979).
\bibitem{EPJArmsHe}V. Lapoux and N.Alamanos, Eur. Phys. J. A {\bf 51}, 91 (2015).
\bibitem{JLM77b} J.P. Jeukenne, A. Lejeune, and  C. Mahaux, Phys. Rev. C   {\bf  16}, 80 (1977).
\bibitem{Pet85} J.S. Petler, M. S. Islam, R.W. Finlay, and F. S. Dietrich,  Phys. Rev. C {\bf  32}, 673 (1985).
\bibitem{Norum82} B. Norum {\it et al.}, Phys. Rev. C {\bf 25}, 1778 (1982).
\bibitem{Oza01} A. Ozawa, T. Suzuki, I. Tanihata, Nucl. Phys. \textbf{A 693}, 32 (2001).
\bibitem{Fab80} E. Fabrici, S. Micheletti, M. Pignanelli, F. G. Resmini, R. De Leo,
 G. D'Erasmo, and A. Pantaleo, Phys. Rev. C {\bf 21}, 844 (1980).
\bibitem{Khan00} E. Khan {\it et al.}, Phys. Lett. B. {\bf  490}, 45 (2000).
\bibitem{Bec06} E. Becheva {\it et al.}, (MUST collaboration), Phys. Rev. Lett. {\textbf 96}, 012501 (2006).
\bibitem{Sly4} E. Chabanat, P Bonche, P. Haensel, J. Meyer, R. Schaeffer, Nucl. Phys \textbf{A 635}, 231 (1998).
\bibitem{DeGo80} J. Decharg\'e and D. Gogny, Phys. Rev. C {\bf 21}, 1568 (1980).
\bibitem{D1S91} J.-F. Berger, M. Girod, and G. Gogny, Comput. Phys. Commun. {\bf 63}, 365 (1991).
\bibitem{Kan11} R. Kanungo {\it et al.}  Phys. Rev. C {\bf 84}, 061304 (R)  (2011). 
\end{references}
\end{document}